\documentclass[a4paper,aps,prd,10pt,preprintnumbers,showpacs,twocolumn,superscriptaddress,nofootinbib]{revtex4-1}

\usepackage[dvips]{graphics}
\usepackage[utf8]{inputenc}
\usepackage[T1]{fontenc}
\usepackage{cmap}

\def\m{\overline{m}}
\def\a{\widetilde{\alpha}}

\def\M{{\cal M}}

\def\Order{{\cal O}}

\ifx\email\undefined
\let\email\ead
\let\affiliation\address
\newenvironment{acknowledgments}{\ack}{}
\ioptwocol
\fi

\begin{document}

\title{Massive particles in the Einstein-Lovelock-anti-de Sitter black hole spacetime}

\author{R. A. Konoplya}\email{roman.konoplya@gmail.com}
\affiliation{Research Centre for Theoretical Physics and Astrophysics, Institute of Physics, Silesian University in Opava, Bezručovo nám. 13, CZ-74601 Opava, Czech Republic}
\affiliation{Peoples Friendship University of Russia (RUDN University), 6 Miklukho-Maklaya Street, Moscow 117198, Russian Federation}

\author{A. Zhidenko}\email{alexander.zhidenko@uni-oldenburg.de}
\affiliation{Institute of Physics, University of Oldenburg, D-26111 Oldenburg, Germany}
\affiliation{Centro de Matemática, Computação e Cognição (CMCC), Universidade Federal do ABC (UFABC),\\ Rua Abolição, CEP: 09210-180, Santo André, SP, Brazil}

\begin{abstract}
An interpretation to the physics of stable geodesics of massive particles in black hole backgrounds has been recently proposed in the context of the AdS/CFT correspondence. It was argued that the existence of stable orbits indicates that the dual state does not thermalize on a thermal time scale and the bulk excitations can be interpreted as metastable states in the dual field theory. Here we study motion of massive particles in the background of the $D$-dimensional asymptotically anti-de Sitter (AdS) black holes in the Einstein-Lovelock theory. We show that, unlike the asymptotically flat case, for any kind of higher curvature Lovelock corrections there appear a stable circular orbit at a distance from the black hole. We find the general analytical expressions for the frequencies of distant circular orbits and radial oscillation frequencies. We show that the corresponding correction has the same power as in the Schwarzschild-AdS case, implying a universal scaling with the temperature for any Lovelock theory.
\end{abstract}

\maketitle

\section{Introduction}

Motion of particles in black-hole spacetimes is usually interesting in the astrophysical context, when the cosmological factors, such as, for example, the positive cosmological constant, are insignificant for motion around compact objects. Unlike the de Sitter case, for asymptotically AdS black holes the interpretation of stable circular orbits has been recently suggested in \cite{Berenstein:2020vlp} within the correspondence between gravitation of $(D+1)$-dimensional asymptotically AdS spacetimes and $D$-dimensional field theory \cite{Maldacena:1997re,Gubser:1998bc}. Since the seminal Tangherlini's paper \cite{Tangherlini:1963bw} it is known that $D>4$ dimensional asymptotically flat analogues of Schwarzschild and Reissner-Nordström solutions do not allow for stable circular orbits. The same is true for higher dimensional asymptotically dS case \cite{Hackmann:2008tu}, but not for asymptotically AdS black holes, where stable circular orbits have been found \cite{Berenstein:2020vlp}. The existence of these stable orbits suggests that the dual state does not thermalize on a thermal time scale and the bulk excitations are interpreted as metastable states in the dual theory \cite{Berenstein:2020vlp}.

At the same time, it is expected that higher-curvature terms added to the Einstein action may represent corrections to the results obtained via AdS/CFT at large coupling, so that the regime of intermediate coupling on the CFT side could be described via analysis of higher dimensional asymptotically AdS black holes in the Einstein-Gauss-Bonnet, Einstein-Lovelock and other higher curvature theories \cite{Solana:2018pbk,Grozdanov:2016fkt,Grozdanov:2016vgg,Konoplya:2017zwo,Andrade:2016rln,Brigante:2008gz}. An essential point when considering black holes in the Einstein-Lovelock theory is dynamical instability of the black hole spacetime which occurs unless the coupling constants are small enough \cite{Gleiser:2005ra,Dotti:2005sq,Takahashi:2011du,Takahashi:2011qda,Konoplya:2017ymp,Konoplya:2017lhs,Konoplya:2020juj,Konoplya:2020bxa,Cuyubamba:2020moe}. The instability of the background at some values of the parameters, means that the black hole metric cannot be used for the dual description under in this regime, so that one must always be constrained by the stable black-hole configurations only.

The null geodesics in the higher dimensional asymptotically flat Einstein-Gauss-Bonnet black hole background were considered in  \cite{Konoplya:2017wot} and in \cite{Hegde:2020yrd} for the four-dimensional asymptotically AdS case. At the same time, the case of geodesics and stability of particle orbits in the $D$-dimensional Einstein-Gauss-Bonnet-AdS spacetime have not been considered so far. Having in mind the above motivations of interpretation of stable circular orbits in AdS for dual description of metastable states on the one side, and of higher curvature corrections for achieving the intermediate coupling regime in the dual theory on the other side, we would like to analyse circular orbits of the $D$-dimensional asymptotically AdS Einstein-Gauss-Bonnet black holes and their Lovelock generalization.
Here we show that the stable orbits are allowed for any asymptotically AdS $D$-dimensional Einstein-Lovelock black hole, which occur for sufficiently large distance from the black hole. The analytical expression for the asymptotic values of the circular-orbit frequency and radial-oscillation frequency of the near circular orbits are derived. The dominant correction to the pure AdS orbit frequency has the same power for any Lovelock theory, implying a universal scaling with the temperature when considering the intermediate coupling in the dual theory.

The paper is organized as follows. In Sec.~\ref{sec:Lovelock} we give the basic information about the black hole metric in the Einstein-Lovelock gravity. Sec.~\ref{sec:AdSas} is devoted to deduction of the asymptotic expansion for the metric function which we will use. In Sec.~\ref{sec:orbits} we derive the equations of motion, the effective potential and the associated frequencies. Finally, in sec.~\ref{sec:conclusions} we summarize the obtained results.

\section{Spherically symmetric black holes in the Lovelock theory}\label{sec:Lovelock}
The Einstein-Lovelock theory is the most general metric theory of gravity keeping the diffeomorphism invariance and yielding second-order equations of motion in an arbitrary number of spacetime dimensions $D$. In $D=4$ it coincides with the Einstein gravity, while in $D=5, 6$ it is represented by the (quadratic in curvature) Einstein-Gauss-Bonnet Lagrangian. Higher $D$ require corrections which are higher than the second order in curvature.

The Einstein-Lovelock theory is given by the Lagrangian density \cite{Lovelock},
\begin{eqnarray}\label{Lagrangian}
  \mathcal{L} &=& -2\Lambda+\sum_{m=1}^{\m}\frac{1}{2^m}\frac{\alpha_m}{m}
  \delta^{\mu_1\nu_1\mu_2\nu_2 \ldots\mu_m\nu_m}_{\lambda_1\sigma_1\lambda_2\sigma_2\ldots\lambda_m\sigma_m}\,\\\nonumber
  &&\times R_{\mu_1\nu_1}^{\phantom{\mu_1\nu_1}\lambda_1\sigma_1} R_{\mu_2\nu_2}^{\phantom{\mu_2\nu_2}\lambda_2\sigma_2} \ldots R_{\mu_m\nu_m}^{\phantom{\mu_m\nu_m}\lambda_m\sigma_m},
\end{eqnarray}
where
$$\delta^{\mu_1\mu_2\ldots\mu_p}_{\nu_1\nu_2\ldots\nu_p}=\det\left(
\begin{array}{cccc}
\delta^{\mu_1}_{\nu_1} & \delta^{\mu_1}_{\nu_2} & \cdots & \delta^{\mu_1}_{\nu_p} \\
\delta^{\mu_2}_{\nu_1} & \delta^{\mu_2}_{\nu_2} & \cdots & \delta^{\mu_2}_{\nu_p} \\
\vdots & \vdots & \ddots & \vdots \\
\delta^{\mu_p}_{\nu_1} & \delta^{\mu_p}_{\nu_2} & \cdots & \delta^{\mu_p}_{\nu_p}
\end{array}
\right)$$
is the generalized totally antisymmetric Kronecker delta, $R_{\mu\nu}^{\phantom{{\mu\nu}}\lambda\sigma}$ is the Riemann tensor, $\alpha_1=1/8\pi G=1$, and $\alpha_2,\alpha_3,\alpha_4,\ldots$ are arbitrary constants of the theory.

The Euler-Lagrange equations have the following form \cite{Kofinas:2007ns}:
\begin{eqnarray}\label{Lovelock}
  0 &=& \Lambda\delta^{\mu}_{\nu}-\sum_{m=1}^{\m}\frac{1}{2^{m+1}}\frac{\alpha_m}{m}
  \delta^{\mu\mu_1\nu_1\mu_2\nu_2 \ldots\mu_m\nu_m}_{\nu\lambda_1\sigma_1\lambda_2\sigma_2\ldots\lambda_m\sigma_m} \\\nonumber
&& \times R_{\mu_1\nu_1}^{\phantom{\mu_1\nu_1}\lambda_1\sigma_1} R_{\mu_2\nu_2}^{\phantom{\mu_2\nu_2}\lambda_2\sigma_2} \ldots R_{\mu_m\nu_m}^{\phantom{\mu_m\nu_m}\lambda_m\sigma_m}\,.
\end{eqnarray}

The antisymmetric tensor is nonzero only when the indices $\mu,\mu_1,\nu_1,\mu_2,\nu_2,\ldots\mu_m,\nu_m$ are all distinct. Thus, the general Lovelock theory is such that $2\m$ is smaller than the number of spacetime dimensions $D$. In particular, in $D=4$ spacetime the Lovelock theorem implies that $\m=1$, which is equivalent to the Einstein theory \cite{Lovelock}. When $D=5$ or $6$, $\m=2$ (Einstein-Gauss-Bonnet theory), implying that the correction to the Einstein action is quadratic in curvature and a new appropriate nonvanishing coupling constant $\alpha_2$ appears.

The generic $D$-dimensional static and maximally symmetric metric is given by the line element
\begin{equation}\label{Lmetric}
  ds^2=-(1-r^2\psi(r))dt^2+\frac{1}{1-r^2\psi(r)}dr^2 + r^2d\Omega_n^2,
\end{equation}
where $d\Omega_n^2$ is a $(n=D-2)$-dimensional sphere, and the function $\psi(r)$ satisfies an algebraic equation \cite{Cai:2003kt}
\begin{equation}\label{MEq}
W[\psi(r)]=\frac{2\M}{r^{D-1}}\,,
\end{equation}
where
\begin{eqnarray}\label{Wpsi}
W[\psi]&=&-\frac{2\Lambda}{(D-1)(D-2)}+\psi+\sum_{m=2}^{\m}\a_m\psi^m,\\\nonumber
\a_m&=&\frac{\alpha_m}{m}\frac{(D-3)!}{(D-2m-1)!}=\frac{\alpha_m}{m}\prod_{p=1}^{2m-2}(D-2-p),
\end{eqnarray}
and the arbitrary constant $\M$ defines the asymptotic mass \cite{Myers:1988ze},
\begin{equation}\label{MADM}
  M=\frac{(D-2)\pi^{D/2-3/2}}{4\Gamma(D/2-1/2)}\M.
\end{equation}

\section{Asymptotically AdS black holes}\label{sec:AdSas}
We further assume that the cosmological constant is negative,
\begin{equation}\label{Ldef}
  \frac{2\Lambda}{(D-1)(D-2)}=-\frac{1}{L^2},
\end{equation}
and the spacetime is asymptotically AdS with the radius $R$ \cite{Konoplya:2017lhs},
\begin{equation}\label{AdSasymp}
  \psi(r\to\infty)=-\frac{1}{R^2}.
\end{equation}

Considering the limit $r\to\infty$ in (\ref{MEq}) we obtain relation between $L$ and $R$,
\begin{equation}\label{RLrel}
  \frac{1}{L^2}=\frac{1}{R^2}+\sum_{m=2}^{\m}(-1)^m\frac{\a_m}{R^{2m}},
\end{equation}
so that for the $\a_m=0$, $L=R$.

Calculating derivative of (\ref{MEq}), we find
\begin{equation}\label{WEq}
W'[\psi(r)]\psi'(r)=-(D-1)\frac{2\M}{r^D}<0\,.
\end{equation}
We notice that for the nonsingular solution $W'[\psi(r)]$ cannot cross zero and is positive at the horizon once we require positiveness of Hawking temperature \cite{Konoplya:2017lhs}.

Finally, we can find the asymptotic expansion for $\psi(r)$,
\begin{equation}\label{psias}
\psi(r)=-\frac{1}{R^2}+\frac{2\M_1}{r^{D-1}}+\frac{2\M_2}{r^{2D-2}}+\Order\left(\frac{1}{r^{3D-3}}\right),
\end{equation}
where
\begin{equation}\label{M1def}
  \M_1=\frac{\M}{W'[-R^{-2}]}=\frac{\M}{1-2\a_2R^{-2}+3\a_3R^{-4}\ldots}>0
\end{equation}
and
\begin{equation}\label{M2def}
  \M_2=-\frac{\M^2 W''[-R^{-2}]}{W'[-R^{-2}]^3}=-\frac{\M_1^3}{\M}\left(2\a_2-\frac{6\a_3}{R^2}\ldots\right).
\end{equation}

While $\M_1$ is always positive, $\M_2$ is zero in the Einstein theory ($\m=1$) and, in the Einstein-Gauss-Bonnet theory ($\m=2$), $\M_2$ and $\a_2$ have opposite signs.

With these expressions at hand we are ready to analyze motion of massive particles in this background.

\section{Particle motion}\label{sec:orbits}
Following \cite{Berenstein:2020vlp}, we consider geodesic motion of a particle of mass $\mu$ in the vicinity of the black hole (\ref{Lmetric}), which is described by the following first-order equations:
\begin{eqnarray}
  e &=& (1-r^2\psi(r))\dot{t} \\
  \ell &=& r^2\dot{\phi} \\
  e^2 &=& \dot{r}^2+V(r),\label{fulleq}
\end{eqnarray}
where $e$ and $\ell$ are integrals of motion (particle's energy and angular momentum, respectively). Dots denote derivatives with respect to the affine parameter. The effective potential has the form
\begin{equation}\label{potential}
  V(r)=(1-r^2\psi(r))\left(\mu^2+\frac{\ell^2}{r^2}\right).
\end{equation}

On the circular orbits $\dot{r}=0$ and $\ddot{r}=0$, and the corresponding radial coordinate satisfies
\begin{equation}\label{circorbits}
  V'(r)=0,
\end{equation}
from which we find the relation between the angular momentum and the coordinate,
\begin{equation}\label{ellr}
  \frac{\ell^2}{\mu^2}=-r^4\frac{2\psi(r)+r\psi'(r)}{2+r^3\psi'(r)}.
\end{equation}

Circular orbits exist ($\ell^2>0$) when the denominator of (\ref{ellr}) is positive. The limiting orbit,
$$2+r^3\psi'(r)=0,$$
corresponds to the circular null geodesic, satisfying (\ref{circorbits}) for $\mu=0$.

The circular orbit is stable, when $V''(r)>0$. In the asymptotically flat spacetime only unstable circular orbits exist for $D>4$ in the Einstein \cite{Tangherlini:1963bw} and Einstein-Gauss-Bonnet theories \cite{Bhawal:1990nh}. Here we show that the circular orbits at a large distance from the black hole are always stabilized in the asymptotically AdS case for any type of the Einstein-Lovelock theory. The second derivative of the effective potential can be reduced to the following form
\begin{eqnarray}\nonumber
V''(r)&=&\ell^2\frac{6-r^4\psi''(r)}{r^4}-\mu^2(2\psi(r)+4r\psi'(r)+r^2\psi''(r)).
\end{eqnarray}
Substituting (\ref{ellr}) into the latter expression and expanding it in powers of $1/r$ we find that
\begin{equation}\label{Vprimeprime}
V''(r)= \frac{8\mu^2}{R^2}+\Order\left(\frac{1}{r^{D-3}}\right).
\end{equation}
Thus, the second derivative of the effective potential is positive for sufficiently distant orbits in AdS.

The corresponding circular-orbit frequency,
\begin{eqnarray}\label{freq}
\omega&=&\frac{d\phi}{dt}=\frac{\ell(1-r^2\psi(r))}{r^2e}=\sqrt{-\psi(r)-\frac{r\psi'(r)}{2}}
\\\nonumber&=&\frac{1}{R}+\M_1 R\frac{D-3}{2r^{D-1}}+\Order\left(\frac{1}{r^{2D-2}}\right).
\end{eqnarray}

Thus, the dominant correction to the pure AdS orbit frequency scales according to the same power law in any Lovelock theory,
$$
\omega R\simeq 1+\M_1 R^2\frac{D-3}{2r^{D-1}}\simeq1+\M_1 R^2\frac{D-3}{2}\left(\frac{R\ell}{\mu}\right)^{-\frac{D-1}{2}}.
$$

Similarly to \cite{Berenstein:2020vlp}, for the binding energy we find the following correction at large $\ell$:
$$e=\int\omega d\ell\simeq\frac{\ell}{R}+\mu-\M_1\left(\frac{R\ell}{\mu}\right)^{-\frac{D-3}{2}}.$$

Finally, considering small oscillations $\delta r$ near the stable orbit $r$, from (\ref{fulleq}) we find
\begin{equation}\label{radialoscillation}
  2e\delta e=\delta \dot{r}^2+\frac{V''(r)}{2} \delta r^2,
\end{equation}
so that the radial-oscillation frequency in units of the coordinate time has the form
\begin{eqnarray}
\omega_r^2&=&\frac{V''(r)}{2\dot{t}^2}=-\frac{1}{2}\Biggl(\psi(r)(8+r^3\psi'(r)-r^4\psi''(r))
\\\nonumber&&\qquad+7r\psi'(r)+2r^4\psi'(r)^2+r^2\psi''(r)\Biggr)
\\\nonumber&=&\frac{4}{R^2}-\M_1\frac{D^2-1}{R^2r^{D-3}}-\M_1\frac{(D-3)(D-5)}{r^{D-1}}+\ldots .
\end{eqnarray}
Hence it follows
\begin{equation}
\omega_r R\simeq2-\M_1\frac{D^2-1}{4}\left(\frac{R\ell}{\mu}\right)^{-\frac{D-3}{2}}.
\end{equation}

\begin{figure}
\centerline{\resizebox{\linewidth}{!}{\includegraphics*{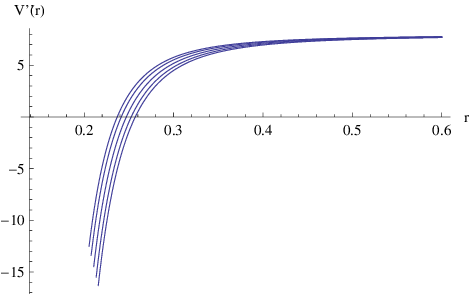}}}
\centerline{\resizebox{\linewidth}{!}{\includegraphics*{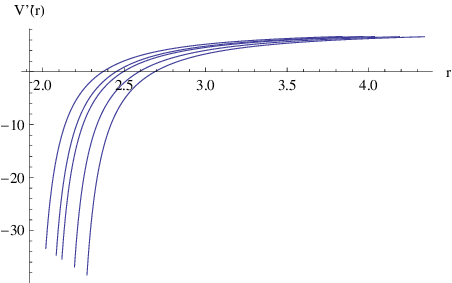}}}
\centerline{\resizebox{\linewidth}{!}{\includegraphics*{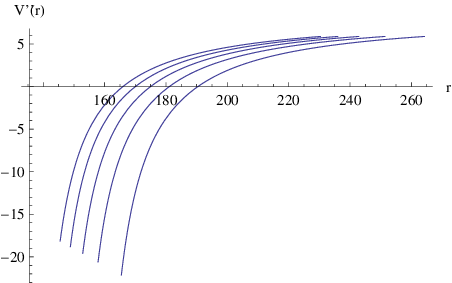}}}
\caption{Second derivative of the effective potential for $D=5$ Einstein-Gauss-Bonnet black holes. From top to bottom:
small black holes ($r_{H} =1/10$, $R=1$), $\a_2 =-9 \cdot 10^{-4}$, $-3 \cdot 10^{-4}$, $5 \cdot 10^{-4}$, $13 \cdot 10^{-4}$, $2 \cdot 10^{-3}$;
intermediate black holes ($r_{H} =R=1$), $\a_2 =-0.05$, $-0.01$, $0.01$, $0.05$, $0.085$;
large black holes ($r_{H} =10$, $R=1$), $\a_2 =-0.11$, $-0.05$, $0.01$, $0.07$, $0.14$.
The first value of $\alpha$ corresponds to the lower threshold of instability and the fifth value is the upper threshold of black-hole instability.
}\label{figs}
\end{figure}

When higher than the second order in curvature corrections are tuned on and the parameters of the metric are not specified, the metric functions of the Einstein-Lovelock black hole cannot be written in a closed analytical form. Nevertheless, one can use the automatic code suggested in \cite{Konoplya:2017lhs} in order to calculate the metric function and its derivatives numerically when the parameters are fixed. Then, the second derivative of the effective potential can be obtained for any value of the radial coordinate.
Our extensive search for various Gauss-Bonnet and Lovelock corrections found no qualitatively new behavior: stable orbits are allowed for any $r>r_{ISCO}$, where $r_{ISCO}$ corresponds to the innermost (marginally stable) circular orbit. This is illustrated in Figs.~\ref{figs}, where the second derivative of the effective potential is shown for small, intermediate, and large (in comparison with the AdS radius $R$) Einstein-Gauss-Bonnet black holes. It is essential that we were limited by the region of stability of the background black hole against small spacetime perturbations.

It is interesting to note that a particular type of instability of the black holes in the Lovelock theory is closely related to the causality of the corresponding space-time. In \cite{Camanho:2014apa} it was shown that, in the Einstein-Gauss-Bonnet theory, the violation of causality can take place for high-energy scattering of gravitons due to the additional structures in the three-point vertex. This happens, in particular, in the vicinity of sufficiently small black holes, where, under some conditions, gravitons experience time advance, which can be used to build a ``time machine'' \cite{Papallo:2015rna}. This problem manifests itself at the classical level when considering linear gravitational perturbations. It was proven in \cite{Takahashi:2010gz} that, once the effective potential in the eikonal limit has negative gap, the black hole is unstable and the instability against perturbations with larger multipole numbers develops faster. Therefore, it was called the eikonal instability \cite{Cuyubamba:2016cug}. At the same time the corresponding black-hole perturbation equations become nonhyperbolic \cite{Reall:2014pwa}, causing a breakdown of the well-posedness of the initial values problem \cite{Papallo:2017qvl}. A nonlinear analysis of the perturbed AdS space-time indicated that such black holes cannot be formed \cite{Deppe:2014oua}, so that the corresponding thermal state in the dual theory does not exist. However, when considering temporal evolution of massive or massless particles of test fields in the vicinity of the Einstein-Lovelock black holes, there is no violation of causality. The corresponding effective potential (\ref{potential}) is positive-definite everywhere outside the black hole, and the initial value problem is well-posed even for the gravitationally unstable black holes. Thus, stable circular orbits are allowed in the whole range of parameters of the Einstein-Lovelock black-hole solutions (\ref{Lmetric}) due to the AdS confinement. Thereby, the above discussed violation of causality does not manifest itself on motion of test particles.

\section{Conclusions}\label{sec:conclusions}
Having in mind possible interpretation within the AdS/CFT correspondence \cite{Berenstein:2020vlp}, we have analyzed circular orbits in the spacetime of the asymptotically AdS black hole in the general Einstein-Lovelock theory. It is shown that stable circular orbits exist independently of the particular choice of the Lovelock corrections or number of spacetime dimensions $D$ due to the confining effect of the AdS space, which leads to the stabilization of the distant circular orbits.

It is interesting to note that a confining behavior was found when studying spectral problems in the asymptotically Gödel spacetimes \cite{Konoplya:2011ig,Konoplya:2011ag}. Therefore, it cannot be excluded that similar stabilization of orbits may occur for black holes immersed in a rotating Universe. This could be an interesting problem for further consideration.

\begin{acknowledgments}
The authors acknowledge the support of the grant 19-03950S of Czech Science Foundation (GAČR).
A.~Z. was supported by the Alexander von Humboldt Foundation, Germany.
\end{acknowledgments}

\end{document}